\documentclass[12pt]{article}

\usepackage{amssymb}
\def\labelmark{}
\def\void{}

{\ifx\void\labelname\def\junk{\end{displaymath}}
\else\def\junk{\end{eqnarray}}\fi\junk\labelmark\def\labelname{}}

\newcommand{\bra}{\begin{array}}
\newcommand{\era}{\end{array}}
\newcommand{\beq}{\begin{equation}}
\newcommand{\eeq}{\end{equation}}
\newcommand{\bqn}{\begin{eqnarray}}
\newcommand{\eqn}{\end{eqnarray}}

\def\BC{\bb C}
\def\_\BC{\bbi C}

\newcommand{\om}{\omega}
 
\newcommand{\la}{\lambda} 
\newcommand{\si}{\sigma}

\newcommand{\ep}{\epsilon}

\newcommand{\be}{\beta} 
\newcommand{\ga}{\gamma} 
\newcommand{\te}{\theta}

\newcommand{\pa}{\partial}
\newcommand{\al}{\alpha}  
  
\newcommand{\de}{\delta}

\newcommand{\st}{\star}
\newcommand{\ti}{\tilde}
\newcommand{\da}{\dagger}

\newcommand{\lb}{\label}
\newcommand{\ov}{\over}
\newcommand{\sq}{\sqrt}
\newcommand{\ev}{\equiv}
\newcommand{\hb}{\hbar}

\newcommand{\PR}[1]{ {\it Phys.~Rev.} {\bf #1}}
\newcommand{\PRL}[1]{ {\it Phys.~Rev.~Lett.} {\bf #1}}

\newcommand{\JMP}[1]{ {\it J. Math.~Phys.} {\bf #1}}

\begin{document}
\begin{titlepage}
\renewcommand{\thefootnote}{\fnsymbol{footnote}}

\begin{flushright}
hep-th/0207269
\end{flushright}

\vspace{13mm}
\begin{center}
{\Large\bf Electrons-Holes on Noncommutative Plane \\
and Hall Effect}

\vspace{8mm}

{ \bf{Ahmed Jellal} 
\footnote{E-mail: {\textsf jellal@gursey.gov.tr }}}\\

\vspace{3mm}
{\em Institut f\"ur Physik, 
Technische Universit\"at Chemnitz\\
D-09107 Chemnitz, Germany}\\

\end{center}

\vspace{5mm}

\begin{abstract}
By considering $N_e$-electrons and $N_h$-holes 
together in uniform
external magnetic and electric fields, we
end up with a total Hall conductivity
$\si_{\rm H}^{\rm tot}$, which
is depending to the difference between 
$N_e$ and $N_h$ and becomes null when
$N_e=N_h$. Dealing with the same 
system but requiring that 
the coordinates of plane are noncommuting,
we obtain a new Hall conductivity 
$\si_{\rm H}^{\rm (tot,nc)}$. 
In the limit $N_e=N_h$,
we find that $\si_{\rm H}^{\rm (tot,nc)}$  is
only noncommutativity parameters $\te_i$-dependent,
which means that theoretically it is possible to have
Hall effect without $B$. Moreover, at
the critical points $\te_e=l^2$ and $\te_h=-l^2$,
we find that $\si_{\rm H}^{\rm (tot,nc)}$
becomes two times the usual Hall conductivity for an
noninteracting mixing system. 

\end{abstract}

\end{titlepage}

\newpage
\section{Introduction}

Two-dimensional (2D) systems of particles under strong 
magnetic fields
exhibiting some beautiful and exciting phenomena like 
fractional quantum Hall effect (FQHE)~\cite{tsui, prange}.
As example of system where FQHE can be appeared, 
one can mention {\it GaAs/AlGaAs} heterostructure~\cite{boebinger} 
and {\it GaAs-Al$_{0.3}$Ga$_{0.7}$}~\cite{chang}. 
As consequence,
one may can think to consider a mixing system 
of electrons and holes moving together in the
plane. As shown in~\cite{chang}, such kind of system 
exists actually in
nature and it is just a doped semiconductor either
with electrons or holes. 

On the other hand, experimentally
a combined system of 2D electron gas (2DEG)
and 2D hole gas (2DHG) can also be formed by 
{\it InAs/AlSb/GaSb}~\cite{lin}. In fact, when 
{\it GaSb} and {\it InAs} are brought together, negative
charge is transferred from {\it GaSb} into
{\it InAs}, creating holes and electrons, respectively,
at the interface of those layers and {\it AlSb} is used as
a barrier. Theoretically, many studies have been carried
on a system formed by separate 2DEG and
2DHG~\cite{girvin,macdonald,demler,mitra}. 
Because it offers the possibility
of forming a gas of excitons aligned along 
the two parallel planes~\cite{lin}. 

Recently with Dayi, we proposed~\cite{jellal1} 
an approach based on noncommutative 
geometry tools~\cite{connes} to describe FQHE 
of a system of electrons. In fact, 
the corresponding filling factor is found to be 
\beq
\lb{dj}
\nu_{\rm DJ} = {\pi\ov 2} \rho ( l^2 - \te )
\eeq
and it is identified to the observed fractional values 
$f=1/3,\ 2/3,\  1/5,\ \cdots$. Also this
approach allowed us to make a link with
composite fermion theory
\cite{jain,heinonen}
of FQHE by setting an effective
magnetic field
\beq
B_{\rm DJ} = {B\ov 1-\te l^2}
\eeq
similar to that felt by composite fermions.

Motived by the above results, we would like to
consider an noninteracting system 
of electrons and holes living on the
plane and in the presence of uniform magnetic field
and electric field. Calculating the 
total Hall conductivity $\si_{\rm H}^{\rm tot}$, 
we find that this latter vanishes 
when the number of electrons $N_e$
coincides with the number of holes $N_h$.
However, another results can be found
when we consider the present system on 
noncommutative plane. Indeed, after deriving 
the corresponding total Hall conductivity
$\si_{\rm H}^{\rm (tot,nc)}$,
we show that in the limit $N_e = N_h$, this
quantity is noncommutativity parameters $\te_i$-dependent
and $B$-independent. On the contrary, it does not vanish 
as noticed in the standard case. Moreover, we present
other discussions and we show that~(\ref{dj})
can be recovered from the present study.
Taking into account of the critical points
$\te_e=l^2$ and $\te_h=-l^2$,
we find that $\si_{\rm H}^{\rm (tot,nc)}$ 
becomes $B$-dependent.

In section 2, we give the energy spectrum 
and the eigenstates of a Hamiltonian describing 
noninteracting electron-hole living on plane and in 
the presence of magnetic and electric fields. 
By using the standard definition
we determine the corresponding total current operator
leading to a total Hall conductivity.
In section 3, we deal with the same problem but in
noncommutative plane and we calculate 
its total Hall conductivity. We discuss and comment 
the obtained results in section~4 and we consider
the critical points of the noncommutative
analysis in section~5.

\section{Hall conductivity on plane}

We would like to study a mixing system, which
may can be the interface between two layers 
of 2DEG and 2DHG or a doped semiconductor. 
In doing on,
let us start by considering a system of an electron
and a hole living together on the plane $(x,y)$ 
and in the presence of an uniform external  
 $\vec B$ and ${\vec E}$ fields. Without interaction, this
system is described by the Hamiltonian
\beq
\label{hami1}
H^{\rm tot} = H_e + H_h
\eeq
as sum of two 
independent Hamiltonian's
corresponding to an electron $H_e$ and
a hole $H_h$, respectively
\beq
\bra{l}
\label{towhami}
H^{e} = {1\over 2m_e}({\vec p}_e+{e\over c}{\vec A}_e)^2
	+eEx_e\\
H^{h} = {1\over 2m_h}({\vec p}_h-{e\over c}{\vec A}_h)^2
	-eEx_h
\era
\eeq
where $({\vec p}_i,{\vec x}_i)$ are electron and hole 
phase spaces, $i=e,h$ denotes electron and hole.
In the symmetric gauges
\beq
\bra{l}
\label{gco}
{\vec A}_e={B\over 2}(-y_e,x_e)\\
{\vec A}_h={B\over 2}(-y_h,x_h)
\era
\eeq
(\ref{hami1}) takes the form
\beq
\bra{l}
\lb{hami2}
H^{\rm tot}=
\frac{1}{2m_e}\left[
\left( p_{x_e} -\frac{eB}{2c} y_e \right)^2 +
\left( p_{y_e} +\frac{eB}{2c} x_e \right)^2 \right] +eEx_e\\ 
\qquad + \; \frac{1}{2m_h}\left[
\left( p_{x_h} +\frac{eB}{2c} y_h \right)^2 +
\left( p_{y_h} -\frac{eB}{2c} x_h \right)^2 \right] -eEx_h.
\era
\eeq
From the last equation, one may can notice that
the sign of charge of the carrier $(e,h)$ is important
and will play a crucial
role in the next. This point will be clear when
we will derive the Hall current and then
the Hall conductivity.

The Hamiltonian $H^{\rm tot}$ can be diagonalised simply by
considering the following operators corresponding
to electron~\cite{jellal1}
\beq
\bra{l}
\lb{eop}
b_e^\da =-2i{p}_{\bar z_e}+{eB\ov 2c}{z}_e+\la_e\\
b_e =2i {p}_{z_e}+{eB\ov 2c}{{\bar z}}_e+\la_e\\
d_e=2i {p}_{z_e}-{eB\ov 2c} {{\bar z}}_e\\
d_e^{\da}=-2i {p}_{\bar z_e}-{eB\ov 2c} {z}_e
\era
\eeq     
and another set related to hole
\beq
\bra{l}
\lb{hop}
b_h^\da =-2i {p}_{\bar z_h}-{eB\ov 2c}{z}_h+\la_h\\
b_h = 2i {p}_{z_h}+{eB\ov 2c} {{\bar z}}_h+\la_h\\
d_h= 2i {p}_{z_h}+{eB\ov 2c}{{\bar z}}_h\\
d_h^{\da}= -2i {p}_{\bar z_h}+{eB\ov 2c}{z}_h.
\era
\eeq  
The $\la_i$'s are fixed to be 
\beq
\bra{l}
\la_e={m_ecE\ov B}\\
\la_h={m_hcE\ov B}.
\era
\eeq
These sets satisfy the commutation relations
\beq
\bra{l}
\lb{cr}
[b_i, b_i^{\da}]= \pm 2m_i\hb\om_i \\

[d_i^{\da},d_i]=\pm  2m_i\hb\om_i\\
\era
\eeq
and other commutators vanish, where plus
refers to electron and minus to hole.
The cyclotron frequencies are given by
\beq
\om_i={eB\over m_ic}.
\eeq
Actually, ${H^{\rm tot}}$ can be expressed
in terms of the above operators as
follows
\beq
\bra{l}
\label{hc}
{H^{\rm tot}}= {1\ov 4m_e}(b_e^{\da}b_e+b_eb_e^{\da})-
{\la_e\ov 2m_e}(d_e^{\da}+d_e)-{\la_e^2\ov 2m_e}\\
\qquad\; 
+ {1\ov 4m_h}(b_h^{\da}b_h+b_hb_h^{\da})-
{\la_h\ov 2m_h}(d_h^{\da}+d_h)-{\la_h^2\ov 2m_h}.
\era
\eeq

From the eigenvalue equation 
\beq
H^{\rm tot}|\Psi^{\rm tot}>= E^{\rm tot}|\Psi^{\rm tot}>
\eeq
we can obtain the 
energy spectrum and eigenstates 
\beq
\lb{hfs}
\bra{l}
\Psi_{(n_e,n_h,\al_e,\al_h)}^{\rm tot}=\Phi_{n_e}\otimes 
\Phi_{n_h}\otimes \phi_{\al_e}\otimes \phi_{\al_h}\ev 
|n_e,n_h,\al_e,\al_h> \\
E_{(n_e,n_h,\al_e,\al_h)}^{\rm tot}={\hb\om_e\ov 2}(2n_e+1)
+ {\hb\om_h\ov 2}(2n_h+1)-
{\hb\la_e\ov m_e}\al_e-{\la_e^2\ov 2m_e}
-
{\hb\la_h\ov m_h}\al_h-{\la_h^2\ov 2m_h}
\era
\eeq
where
\beq
\bra{l}
\Phi_{n_i} = 
{1\ov \sqrt{(2m_i\hb{\om}_i)^{n_i} n_i!}}
({b}_i^{\da})^{n_i}|0>\\
\phi_{\al_i} = e^{i(\al_i y_i+{m_i{\om_i}\ov 2\hb}x_iy_i)}.
\era
\eeq
$n_e,n_h=0,1,2\cdots$ and $\al_e,\al_h \in  \mathbb{R}$.
$\otimes $ denotes the direct  product.

At this stage, we would like to determine the 
corresponding total Hall conductivity 
$\si_{\rm H}^{\rm tot}$ in order to
have some informations about the behaviour
of the mixing system. 
To derive this physical quantity,
one can use directly the definition of
the related total current operator 
${\vec{J}}^{\rm tot}$,
such as
\beq
\lb{nco}
{{\vec{J}}}^{\rm tot} ={{\vec{J}}}_e + {{\vec{J}}}_h
\eeq
where ${{\vec{J}}}_e$ and ${{\vec{J}}}_h$
are defined to be
\beq
\bra{l}
{{\vec{J}}}_e=-{e\rho_e\ov m_e} ({\vec p }_e+{e\ov c}{\vec A}_e)\\
{{\vec{J}}}_h={e\rho_h\ov m_h} ({\vec p }_h-{e\ov c}{\vec A}_h)
\era
\eeq
and $\rho_e$ and $\rho_h$ are, respectively,
electron and hole densities. Let us emphasis here that
the sign of charge of different particles
is taken account.
Moreover,
the expectation value of ${{\vec{J}}}^{\rm tot}$ 
can be calculated with respect to the eigenstates 
$|n_e,n_h,\al_e,\al_h>$ (\ref{hfs}). Therefore, 
we obtain
\beq
\lb{ncco}
\bra{l}
< {J}_x^{\rm tot} >=0 \\
<{J}_y^{\rm tot}>={ec\ov B}\Big(\rho_e - \rho_h \Big)E.
\era
\eeq
The second equation determines 
the so-called Hall conductivity and then 
leads us to have a total
$\si_{\rm H}^{\rm tot}$ as
\beq
\lb{hc}
\si_{\rm H}^{\rm tot}={ec\ov B}\Big(\rho_e - \rho_h \Big)
\eeq
which is actually sum of two contributions coming
from electrons and holes, respectively
\beq
\bra{l}
\si_{\rm H}^{\rm e}={ec\rho_e\ov B}\\
\si_{\rm H}^{\rm h}=-{ec\rho_h\ov B}.
\era 
\eeq
At this level, let us notice two 
remarks. First, one may can see
that once we have $\rho_h = \rho_e$,  
$\si_{\rm H}^{\rm tot}$ becomes null.
Then in this case the mixing system is
behaving like an insulator.
Second, if one drops $\si_{\rm H}^{\rm tot}$ as function
of the magnetic field $B$, we end up with
straight line. However, the experiment observation
claimed that at strong $B$, we have plateaus. 
Consequently, when the straight line meets 
the plateaus, it is equivalent to have
\beq 
\lb{iqhe}
{ec\ov B}\Big(\rho_e - \rho_h \Big) = 
\nu^{\rm tot} {e^2\ov h}
\eeq
where $\nu^{\rm tot}$ is the total filling factor, 
which is characterising quantum Hall effect. This equation
leads us to have 
\beq
\nu^{\rm tot} = { N_e - N_h \ov N_{\Phi_B}}
\eeq
it is nothing but the
definition of $\nu^{\rm tot}$, i.e. 
the ratio between number of particles
ensuring the conductivity 
and number of quantum flux 
$N_{\Phi_B}={\Phi_B\ov \Phi_0}$, 
where $\Phi_0 = hc/e$ is unit
of flux.

\section{Hall conductivity on noncommutative plane}

The results derived in the last section can be 
generalised in terms of noncommutative
geometry~\cite{connes} and lead us to have more
informations, especially for total Hall
conductivity. To clarify these points,
let us start by demanding that the coordinates 
of the plane 
are noncommuting, which means that the
spacial commutator is now broken such that 
\beq
[x^{i},x^{j}]=i\te^{ij}
\label{nccoo}
\eeq
where $\te^{ij}=\ep^{ij}\te$ is the 
noncommutativity parameter
and $\ep^{12}=-\ep^{21}=1$. Basically, 
we are forced in this case
to replace $fg(x)=f(x)g(x)$ by the relation
\beq
f(x) \st g(x)=\exp[{i\over 2}\te^{ij}
\pa_{x^{i}}\pa_{y^{j}}]f(x)g(y)|_{x=y}
\label{2}
\eeq
where $f$ and $g$ are two arbitrary functions, supposed to be 
infinitely differentiable.
As consequence, now we are going to deal with 
quantum mechanics by considering the
following algebra
\beq
\bra{lll}
\lb{deqm}
[x^{i},x^{j}]=i\te^{ij}\\

[p^{i},x^{j}]=-i\de^{ij}\\

[p^{i},p^{j}]=0.
\era
\eeq

At this level, one may can use the above receipt
to deform the electron's and hole's
phase space independently~\cite{dayi}. Thus,
instead of using~(\ref{nccoo}), we consider
two different relations, namely one with
respect to electrons
\beq
\bra{l}
\lb{defel}
[x_e,y_e]=i\te_e
\era
\eeq
and another one related to holes
\beq
\bra{l}
\lb{defho}
[x_h,y_h]=i\te_h
\era
\eeq
where $\te_i$'s are real parameter, and 
the usual canonical quantization similar to
the two last equations given in~(\ref{deqm}).

Actually, we can write down
the noncommutative version of the Hamiltonian~(\ref{hami1}). 
In doing on, let us notice $H^{\rm tot}$
acts on an arbitrary function $\Psi(\vec{r},t)$ as
\beq
\bra{l}
\lb{nhami}
H^{\rm tot} \st \Psi (\vec{r},t) = H^{\rm (tot,nc)} \Psi (\vec{r},t)
\era
\eeq
which implies that $H^{\rm (tot,nc)}$ is
\beq
\bra{l}
\lb{nh}
{H}^{\rm (tot,nc)} =
\frac{1}{2m_e}\left[
\left( \ga_e {p}_{x_e} -\frac{eB}{2c} {y}_e \right)^2 +
\left(\ga_e {p}_{y_e} +\frac{eB}{2c} {x}_e \right)^2 \right]
+eE({x}_e-{\te_e\ov 2\hb}{p}_{y_e})\\
\qquad\qquad
+ \frac{1}{2m_h}\left[
\left( \ga_h {p}_{x_h} +\frac{eB}{2c} {y}_h \right)^2 +
\left(\ga_h {p}_{y_h} -\frac{eB}{2c} {x}_h \right)^2 \right]
-eE({x}_h -{\te_h\ov 2\hb} {p}_{y_h})
\era
\eeq
where the $\ga_{i}$'s are defined to be
\beq
\bra{l}
\ga_{e} = 1- \te_e l^{-2}\\
\ga_{h} = 1+ \te_h l^{-2}
\era
\eeq
and $l=2l_0$,  $l_0=\sq{\hb c\ov eB}$ is  
the magnetic length. Note in passing 
that~(\ref{nh}) is also a sum of two
noncommutative parts of $H_e$ and
$H_h$:
\beq
H^{\rm (tot,nc)} = H_e^{\rm nc}+ H_h^{\rm nc}.
\eeq

Now, one can process as before to diagonalise
the noncommutative Hamiltonian $H^{\rm (tot,nc)}$. 
For that, let us define the following operators
for electron
\beq   
\lb{nob}
\bra{l}
{\ti b}_e^{\da}=-2i \ga_e{{ p}}_{\bar z_e}+{eB\ov 2c} {z_e}+\la_{e-} \\
{\ti b}_e=2i \ga_e {{ p}}_{z_e}+{eB\ov 2c} {\bar z_e}+\la_{e-}
\era
\eeq
and 
\beq
\lb{nod}   
\bra{l}
{\ti d}_e=2i \ga_e {{p}}_{z_e}-{eB\ov 2c} {{\bar z_e}}\\
{\ti d}_e^{\da}=-2i  \ga_e {{p}}_{\bar z_e}-{eB\ov 2c}{z_e}.
\era
\eeq
Also in similar way one can define another set
of operators for hole
\beq   
\lb{nhob}
\bra{l}
{\ti b}_h^{\da}= -2i \ga_h {{p}}_{\bar z_h}-{eB\ov 2c} {z_h}+\la_{h-} \\
{\ti b}_h= 2i \ga_h {{p}}_{z_h}-{eB\ov 2c} {\bar z_h}+\la_{h-}
\era
\eeq
and 
\beq
\lb{nhod}   
\bra{l}
{\ti d}_h= 2i \ga_h {{p}}_{z_h}+{eB\ov 2c} {{\bar z_h}}\\
{\ti d}_h^{\da}= -2i  \ga_h {{p}}_{\bar z_h}+{eB\ov 2c}{z_h}.
\era
\eeq
The sets of operators $({\ti b}_i,{\ti b}_i^\da )$ and
$({\ti d}_i,{\ti d}_i^\da )$ commute with each other. Moreover,
they verify the commutation relations 
\beq   
\bra{l}
\lb{ncr}
[{\ti b}_i ,{\ti b}_i^{\da}]= {\pm} 2m_i\hb{\ti\om_i} \\
 
[{\ti d}_i^{\da},{\ti d}_i]= \pm 2m_i\hb{\ti\om_i}\\
\era
\eeq
where the ${\ti\om}_i$'s and the $\la_{\pm i}$'s 
are given by
\beq
\bra{l}
{\ti\om}_e=\ga_e\om_e \\
{\ti\om}_h=\ga_h\om_h \\
\la_{e\pm}=\la_e \pm{em_eE\te_e\ov 4\ga_e\hb}\\ 
\la_{h\pm}=\la_h \mp{em_hE\te_h\ov 4\ga_h\hb}.
\era
\eeq
To make these equations holding and
for further analysis, we assume that
the conditions  $\theta_e\ne l^2$ 
and $\theta_h\ne -l^2$ are satisfied.
We will come back to this assumption in 
the last section and discuss
its consequence. 
In terms of the above creation and annihilation
operators, 
the Hamiltonian ${H}^{\rm (tot,nc)}$ takes the
form
\beq
\bra{l}
\label{nh1}
{H}^{\rm (tot,nc)}= {1\ov 4m_e}({\ti b}_e^{\da}{\ti b}_e+
{\ti b}_e{\ti b}_e^{\da})
-{\la_{e+}\ov 2m_e}
({\ti d}_e^{\da}+{\ti d}_e)-{\la_{e-}^2\ov 2m_e}\\
\qquad\qquad
+ {1\ov 4m_h}({\ti b}_h^{\da}{\ti b}_h+
{\ti b}_h{\ti b}_h^{\da})
-{\la_{h+}\ov 2m_h}
({\ti d}_h^{\da}+{\ti d}_h)-{\la_{h-}^2\ov 2m_h}.
\era
\eeq

As before, we can solve the eigenvalue equation
\beq
{H}^{\rm (tot,nc)}|\Psi^{\rm (tot,nc)}> = 
E^{\rm (tot,nc)}|\Psi^{\rm (tot,nc)}>
\eeq
to get the eigenstates: 
\beq   
\bra{l}
\lb{nef}   
\Psi_{(n_e,n_h,\al_e,\al_h,\te_e,\te_h )}^{\rm (tot,nc)}=
\Psi_{(n_e,\al_e,\te_e)}^{\rm nc} \otimes
\Psi_{(n_h,\al_h,\te_h)}^{\rm nc}
\ev|n_e,n_h,\al_e,\al_h,\te_e,\te_h>\\
\era
\eeq
where
\beq
\Psi_{(n_i,\al_i,\te_i)}^{\rm nc}=
{1\ov \sqrt{(2m_i\hb{\ti\om}_i)^{n_i}}}
e^{i(\al_i y_i+{m_e{\ti\om_i}\ov 2\hb}x_iy_i)}
({\ti b}_i^{\da})^{n_i}|0>
\eeq
and the corresponding eigenvalues:  
\beq
\bra{l}
\lb{nev}  
E_{(n_e,n_h,\al_e,\al_h,\te_e,\te_h)}^{\rm (tot,nc)}=
E_{(n_e,\al_e,\te_e)}^{\rm nc} + 
E_{(n_h,\al_h,\te_h)}^{\rm nc}\\
\era
\eeq
where 
\beq
\bra{l}
E^{\rm nc}_{(n_e,\al_e,\te_e)}=
{\hb{\ti\om}_e\ov 2}(2n_e+1)-
{\hb\ga_e\la_{e+}\ov m_e}\al_e-{m_e\ov 2}\la_{e-}^2 \\
E^{\rm nc}_{(n_h,\al_h,\te_h)}=
{\hb{\ti\om}_h\ov 2}(2n_h+1)-
{\hb\ga_h\la_{h+}\ov m_h}\al_h-{m_h\ov 2}\la_{h-}^2
\era
\eeq
with $n_i=0,1,2...$ and $\al_i\in \mathbb{R}$.

Here one may can ask about the corresponding
conductivity to see the difference with
the commutative case. Actually this quantity 
resulting from the Hamiltonian 
${H}^{\rm (tot,nc)}$ is determined by
defining the total current operator 
${\vec{J}}^{\rm (tot,nc)}$ 
on noncommutative plane as
\beq
\lb{nco}
{{\vec{J}}}^{\rm (tot,nc)} = -
{e\rho_e \ga_e\ov m_e} ( \gamma_e {\vec p}_e+
{e\ov c}{\vec A}_e+{\vec a}_e)
+ {e\rho_h\ga_h\ov m_h} ( \gamma_h {\vec p}_h-
{e\ov c}{\vec A}_h+{\vec a}_h)
\eeq
where the ${\vec a}_i$ vectors are
\beq
\bra{l}
{\vec a}_e=(0,-{m_e eE\te_e\ov 2\hb\ga_e})\\
{\vec a}_h=(0,{m_h eE\te_h\ov 2\hb\ga_h}).
\era
\eeq
Its expectation value are calculated with respect 
to the eigenstates $|n_e,n_h,\al_e,\al_h,\te_e,\te_h>$
(\ref{nef}) and it found to be
\beq
\lb{ncco}
\bra{l}
<{J}_x^{\rm (tot,nc)}> =0\\
<{J}_y^{\rm (tot,nc)}> ={ec\ov B}\Big(\rho_e\ga_e - \rho_h\ga_h \Big)E.
\era
\eeq
Therefore, the total Hall conductivity on 
noncommutative plane of electrons and holes,
denoted by $\si_{\rm H}^{\rm (tot,nc)}$, is
\beq
\lb{nhc}
\si_{\rm H}^{\rm (tot,nc)}= 
{ec\ov B}\Big(\rho_e\ga_e - \rho_h\ga_h\Big).
\eeq

To close this section, let us notice that the  
commutative analysis is recovered if the 
noncommutativity parameters $\te_i$ are 
switched off.

\section{Discussions}

In this section,
we would like to comment the obtained result~(\ref{nhc})
and then discuss its consequence by considering some
special cases either related to different
densities or noncommutativity parameters. Indeed,
let us distinguish some relevant cases:\\

\noindent {\textbf{First}}: ~(\ref{nhc}) can be written as follows
\beq
\si_{\rm H}^{\rm (tot,nc)}= \si_{\rm H}^{\rm tot}
+
{\pi\ov 2}\Big(\te_h\rho_h - \te_e\rho_e \Big) {e^2\ov h}.
\eeq
Then, the second term can be interpreted as
a quantum correction to standard Hall effect,
which is noncommutativity parameters dependent. 
It is obvious to see that 
\beq
\si_{\rm H}^{\rm (tot,nc)}|_{\te_e=\te_h=0}
= \si_{\rm H}^{\rm tot}
\eeq
once we have $\te_e=\te_h=0$.
\\

\noindent {\textbf{Second}} : 
Let us make contact with quantum Hall effect by 
defining the corresponding total effective filling factor 
\beq
\lb{rff}
\nu_{\rm eff}^{\rm tot}={\pi \ov 2 }l^2 
\Big(\rho_e\ga_e - \rho_h\ga_h \Big)
\eeq
which implies that
\beq
\lb{qhenhc}
\si_{\rm H}^{\rm (tot,nc)}= 
\nu_{\rm eff}^{\rm tot} {e^2\ov h}.
\eeq
It is similar to the ordinary relation for Hall conductivity
$\si_{\rm H}^{\rm tot}=  \nu^{\rm tot} {e^2\ov h}$.
Then, one may can interpret~(\ref{nhc}) as a result
of QHE by setting an effective density and magnetic field,
such that their ratio is 
\beq
{\rho_{\rm eff}\ov B_{\rm eff}}= 
\rho_e{\ga_e\ov B} - \rho_h{\ga_h\ov B} 
\eeq 
and leading to
\beq
\si_{\rm H}^{\rm (tot,nc)}= 
{ec\rho_{\rm eff} \ov B_{\rm eff}}
= \nu_{\rm eff}^{\rm tot} {e^2\ov h}
\eeq

\noindent {\textbf{Third}} :
One can  see easily that~(\ref{nhc}) vanishes when
the relation is satisfied
\beq
{\ga_e\ov \ga_h} = {\rho_h\ov \rho_e} 
\eeq  
which is independent to the relationship
between different densities. It is on the
contrary with the standard case where
the total Hall conductivity is found to be
zero whether $N_e=N_h$. \\

\noindent {\textbf{Fourth}} : If we assume that
\beq
{\rho_e\ga_e} = \be {\rho_h \ga_h} 
\eeq 
for any $\be$, then~(\ref{nhc}) gives us
\beq
\si_{\rm H}^{\rm (tot,nc)}= (\be - 1) {ec\ov B}\rho_h\ga_h.
\eeq 
It is $(\be - 1)$ times of what we 
 obtained~\cite{jellal1} for
one system
and coincides with our former results when $\be=2$.\\

\noindent {\textbf{Fifth}} : If we demand that the numbers of
electrons and holes are the same, namely
$\rho_h=\rho_e=\rho$, then~(\ref{nhc}) becomes
\beq
\lb{eqnhc}
\si_{\rm H}^{\rm (tot,nc)}|_{\rho_h=\rho_e=\rho}= 
- {\pi\rho\ov 2} (\te_e + \te_h) {e^2\ov h}
\eeq
and therefore
\beq
\lb{rff}
\nu_{\rm eff}^{\rm tot}|_{\rho_h=\rho_e=\rho}=-
{\pi\rho\ov 2} (\te_e + \te_h). 
\eeq
As consequence, if $\te_e = \te_h=\te $, we find
\beq
\lb{tnhc}
\si_{\rm H}^{\rm (tot,nc)}|_{\rho_h=\rho_e=\rho}^{\te_e = \te_h=\te}= 
- {\pi\rho} \te {e^2\ov h}
\eeq
and obviously~(\ref{eqnhc}) becomes zero
when $\te_e = -\te_h $.
These results are
completely different from standard case where
$\si_{\rm H}^{\rm tot}$ is found to be null 
at the present limit $(\rho_e=\rho_h=\rho)$.
These equations tell us about the possibility to
have theoretically Hall effect without magnetic
field just by controlling the noncommutativity
parameter $\te$ for fixed densities. 
As consequence, one may can ask
about the experiment realisation of such kind
of result, otherwise is it possible to 
prove~(\ref{eqnhc},\ref{tnhc}) experimentally? \\

\section{Critical points}

Another result can be obtained by taking
into account of critical points. In fact,
following the above analysis we have to
consider two different critical values of 
$\te_i$, namely $\te_e = l^2$ and 
$\te_h = -l^2$. Then, let us see what is the
consequence of these particular values
of $\te_i$ on the above results of the
noncommutative study.\\

\noindent {\textbf{(i)}} $\te_e = l^2$:  
The relation~(\ref{defel}) reads as
\beq
[x_e^i, x_e^j]= il^2{\de^{ij}}.
\eeq
In this case the whole Hamiltonian becomes
\beq
H^{\rm (tot,nc)}|_{\te_e= l^2}=
{m\om_e^2\ov 4}\Big(x_e^2 + y_e^2\Big) + 
{2Ec\ov B} \Big({eB\ov 2c}x_e - p_{y_e}\Big)+
H_h^{\rm nc}.
\eeq
Using the same techniques 
as before, we find that~(\ref{nhc}) 
now reads as 
\beq
\lb{hlll}
\si_{\rm H}^{\rm (tot,nc)}|_{\te_e= l^2}= 
{2 ec \rho_e\ov B} -{ec \rho_h\ov B}\ga_h, 
\eeq
which is $\te_h$-dependent. Here also one may can
report the above discussion as concerning
noncommutativity parameter.\\

\noindent {\textbf{(ii)}}  $\te_h = -l^2$: 
A similar result as before can be found
just by switching the indices's. Indeed, now
the deformed hole space is characterised by
\beq
[x_h^i, x_h^j]= - il^2{\de^{ij}}
\eeq
instead of~(\ref{defho}). 
Then, the corresponding Hall 
conductivity is 
\beq
\si_{\rm H}^{\rm (tot,nc)}|_{\te_h= -l^2}= 
{ec \rho_e\ov B}\ga_e - {2 ec \rho_h\ov B} 
\eeq
which is now $\te_e$-dependent.\\

\noindent {\textbf{(iii)}} $\te_e = l^2 = -\te_h$:
(\ref{nhami}) reduces to the
following one
\beq
\bra{l}
H^{\rm (tot,nc)}|_{\te_e= l^2 = -\te_e}=
{m\om_e^2\ov 4}\Big(x_e^2 + y_e^2\Big) + 
{2Ec\ov B} \Big({eB\ov 2c}x_e - p_{y_e}\Big)\\
\qquad \qquad \qquad \qquad   
+
{m\om_h^2\ov 4}\Big(x_h^2 + y_h^2\Big) -
{2Ec\ov B} \Big({eB\ov 2c}x_h - p_{y_h}\Big).
\era
\eeq
Therefore, we end up with 
\beq
\si_{\rm H}^{\rm (tot,nc)}|_{\te_e= l^2= -\te_h}= 
2{ec\ov B}\Big(\rho_e - \rho_h\Big). 
\eeq
Note that the obtained relation is actually 
two times the standard result $\si_{\rm H}^{\rm tot}$
and becomes zero once we have $N_e=N_h$.

\section{Conclusion}

By considering a system of the electrons and holes 
living together on noncommutative plane and
in the presence of magnetic and electric fields,
we ended up with the corresponding total Hall 
conductivity. This
latter has some particularities in sense that
it is noncommutativity parameters dependent
and showed some other behaviour than the 
usual one. In fact, on the contrary to
standard case, it does not vanish whether we have
$N_e=N_h$ and then became magnetic field
independent. Therefore, this result suggested
the possibility to have Hall effect without
magnetic field. Moreover,
by taking into account of different
critical points, we got the usual
Hall conductivity for a mixing system times 
a factor two.

One may can ask about the possibility to
deal with an interacting system of electrons and
holes in the framework of noncommutative geometry.
This problem and related
questions will be considered in 
the feature.

\section*{Acknowledgments}

The author would like to thank \"O.F. Dayi
for fruitful discussion and suggestion. He is also
thankful to P. Cain, M.L. Ndawana and R.A. R\"omer 
for discussion. A Special thanks from AJ to 
E.H.~Saidi for his comment.

\end{document}